\documentclass[thmsa,a4paper]{article}
\usepackage{sw20lart}

\input{tcilatex}
\begin{document}

\author{S. Manoff \\
\textit{Bulgarian Academy of Sciences}\\
\textit{Institute for Nuclear Research and Nuclear Energy}\\
\textit{Department of Theoretical Physics}\\
\textit{Blvd. Tzarigradsko Chaussee 72}\\
\textit{1784 Sofia - Bulgaria}}
\title{\textsc{Local criteria for the existence of an accelerated frame of reference%
}}
\date{\textit{E-mail address: smanov@inrne.bas.bg}}
\maketitle

\begin{abstract}
A local criteria for the existence of an accelerated frame of reference is
found. An accelerated frame of reference could exist in all regions where a
non-null (non-isotropic) vector field does not degenerate in a null
(isotropic) vector field.
\end{abstract}

\section{Introduction}

One of the important problem arising in considerations of a frame of
reference in spaces with affine connections and metrics is finding
conditions for the existence of a frame of reference and especially for the
existence of an accelerated frame of reference \cite{Hehl}.

Every frame of reference \cite{Manoff-1} is described by a set of the
following geometrical objects:

(a) A differentiable manifold $M$ ($dimM=n$) considered as a model of space
or space-time.

(b) A non-null (non-isotropic) contravariant vector field $u$ [$u\in T(M)$, $%
g(u,u)=e\neq 0$].

(c) A tangent subspace (hypersurface) $T_x^{\perp u}(M)$, orthogonal to $u$
at every point $x\in M$, where $u$ is defined. This hypersurface (subspace)
is determined by the vector field $u$ and its corresponding projective
metrics $h_u=g-\frac 1e\cdot g(u)\otimes g(u)$ and $h^u=\overline{g}-\frac
1e\cdot u\otimes u$ as well as by the orthogonal to $u$ contravariant vector
fields $\xi _{(a)\perp }$ [$\xi _{(a)\perp }\in T(M)$, $a=1,\ldots ,n-1$, $%
g(u,\xi _{(a)\perp })=0$].

(d) (Contravariant) affine connection $\nabla =\Gamma $. The affine
connection $\Gamma $ is related to the covariant differential operator $%
\nabla _u$ along $u$, determining the transport of tensor fields along the
vector field $u$.

For the existence of an accelerated frame of reference, the hypersurfaces
determined by a given vector field $u$ at different points of $M$ should not
intersect each other. The vector field $u$ is usually interpreted as the
velocity vector field of observers which trajectories belong to a congruence
of lines [set of non-intersecting lines (curves)] with (at least one)
parameter, related to the proper time of the observer. Then the trajectories
are called world lines of the observers. Every hypersurface, orthogonal to $%
u $, under this interpretation has constant proper time. The hypersurfaces
at different points, where $u$ is defined, have different constant proper
times.

For one observer [one trajectory (curve) with tangent vector $u$ at every of
its points], it could happen that the hypersurfaces, orthogonal to $u$ at
different points of the curve, intersect each other. If the trajectory of
the observer is interpreted as its world line (the parameter of the curve is
related to its proper time) then the intersection points will not have a
unique proper time. One and the same intersection point will have different
proper times from the point of view of the different hypersurfaces to which
it belongs. This ambiguity of the definition of the proper time at the
intersecting points could be considered and interpreted in two different
ways:

(a) The intersection points of hypersurfaces with different (constant)
proper times are boundary points of the region where an accelerated frame of
reference could exist. The reason for such assumption is that for $dimM=4$ a
hypersurface, orthogonal to an observer's world line, appears as the $3$%
-dimensional real world at a given proper time of the observer. An
intersection point is an event belonging to the $3$-dimensional world at
different moments of the proper time of the observer. The event cannot be
distinguished in the time from point of view of the observer. Its existence
could not be uniquely described by the use of a frame of reference since it
is not determined uniquely in the time scale of the frame of reference.

(b) The intersection points of two hypersurfaces (with different constant
proper times of an observer to which world line they are orthogonal) are
events which existence could not be described by a time interval of the
proper time of the observer. The notion of time interval for intersection
points is meaningless: two events with different proper times on the world
line of the observer would both appear simultaneously with an event at an
intersection point of the two hypersurfaces with the corresponding proper
time. This fact requires a closer investigation related to the conditions
for existence of intersection points.

\section{Local conditions for the existence of an intersection point}

1. Let us consider two-parametric curves from the congruence $x^i(\tau
,\lambda )$ with the tangent vectors $u$ and $\xi _{\perp }$, where $g(u,\xi
_{\perp })=0$. Let the parameters of the curves be $\tau $ and $\lambda $
correspondingly. Then the tangent vectors $u$ and $\xi _{\perp }$ could be
written in a co-ordinate basis in the forms 
\[
u=\frac d{d\tau }=\frac{dx^i}{d\tau }\cdot \partial _i=u^i\cdot \partial _i%
\text{ \thinspace \thinspace \thinspace \thinspace \thinspace \thinspace
,\thinspace \thinspace \thinspace \thinspace \thinspace \thinspace
\thinspace \thinspace \thinspace \thinspace \thinspace \thinspace \thinspace
\thinspace }\xi _{\perp }=\frac d{d\lambda }=\frac{dx^j}{d\lambda }\cdot
\partial _j=\xi _{\perp }^j\cdot \partial _j\,\,\,\,\text{.} 
\]

If a frame of reference is well defined in $M$ then there should be no
intersection points of the subspaces (hypersurfaces), orthogonal to $u$ at
different points of the curve with tangent vector $u$.

Let us investigate the opposite case when two hypersurfaces, orthogonal to $%
u $ at different points of the curve $x^i(\tau ,\lambda =const.)$, intersect
each other and violate in this way the unique definition of frame of
reference.

Let two two-parametric sets of curves $x^i(\tau ,\lambda )$ be given with
the curves $x^i(\tau =\tau _0=const.,\lambda )$ and $x^i(\tau ,\lambda
=\lambda _0=const.)$. Let us assume that a curve $x^i(\tau =\tau
_0=const.,\lambda )$ intersects a curve $x^i(\tau =\tau _0+d\tau ,\lambda )$
at a point $B$ with co-ordinates $\overline{x}^i=x^i(\tau ,\lambda )$. From
a point $A$ with co-ordinates $x^i(\tau _0,\lambda _0)$ at the curve $%
x^i(\tau =\tau _0=const.,\lambda )$ the point $B$, lying along the same
curve, letl have the co-ordinates $\overline{x}^i=x^i(\tau _0,\lambda
_0+d\lambda )$. From a point $C$ with co-ordinates $x^i(\tau _0+d\tau
,\lambda _0)$ at the curve $x^i=x^i(\tau _0+d\tau ,\lambda )$, the point $B$%
, lying as intersecting point also at this curve, let have the co-ordinates $%
\overline{x}^i=x^i(\tau +d\tau ,\lambda _0+k\cdot d\lambda )$.

If the co-ordinates $\overline{x}^i$ from both the points $A$ and $C$ of the
curve $x^i(\tau ,\lambda _0)$ are identical (indistinguishable) then 
\[
\overline{x}^i=x^i(\tau _0,\lambda _0+d\lambda )=x^i(\tau +d\tau ,\lambda
_0+k\cdot d\lambda )\,\,\,\,\,\text{.} 
\]

The co-ordinates of the point $B$ could be expressed by the co-ordinates of
the point $A$ or of point $C$ by the use of the exponent of the ordinary
differential operator \cite{Schutz}. On the one side, $\overline{x}^i$ could
be expressed by the co-ordinates of point $C$ and then the co-ordinates of
point $C$ could be expressed by the co-ordinates of the point $A$. 
\begin{eqnarray*}
\stackunder{ACB}{\overline{x}^i} &=&x^i(\tau +d\tau ,\lambda _0+k\cdot
d\lambda )= \\
&=&\{(exp[k\cdot d\lambda \cdot \frac d{d\lambda }])x^i\}_{(\tau _0+d\tau
,\lambda _0)}= \\
&&\left\{ (exp[k\cdot d\lambda \cdot \frac d{d\lambda }])\circ (exp[d\tau
\cdot \frac d{d\tau }])x^i\right\} _{(\tau _0,\lambda _0)}\,\,\,\,\,\,\text{.%
}
\end{eqnarray*}

On the other side, $\overline{x}^i$ could be expressed directly by the
co-ordinates of the point $A$%
\begin{eqnarray*}
\stackunder{AB}{\overline{x}^i} &=&x^i(\tau _0,\lambda _0+d\lambda )= \\
&=&\{(exp[d\lambda \cdot \frac d{d\lambda }])x^i\}_{(\tau _0,\lambda
_0)}\,\,\,\,\,\,\text{.}
\end{eqnarray*}

Since $\stackunder{ACB}{\overline{x}}^i$ and $\stackunder{AB}{\overline{x}}%
^i $ should be identical with the co-ordinates of the point $B$ from point
of view of observers at point $A$ and $C$, we have the condition 
\[
\stackunder{ACB}{\overline{x}^i}=\,\stackunder{AB}{\overline{x}^i}\,\,\,\,%
\text{.} 
\]

Then 
\[
\left\{ (exp[k\cdot d\lambda \cdot \frac d{d\lambda }])\circ (exp[d\tau
\cdot \frac d{d\tau }])x^i\right\} _{(\tau _0,\lambda _0)}=\{(exp[d\lambda
\cdot \frac d{d\lambda }])x^i\}_{(\tau _0,\lambda _0)}\text{ \thinspace
\thinspace .} 
\]

The last expression could be written more explicitly up to the second order
of $d\tau $ and $d\lambda $ as 
\[
\{[1+k\cdot d\lambda \cdot \frac d{d\lambda }+\frac 1{2!}\cdot k^2\cdot
d\lambda ^2\cdot \frac{d^2}{d\lambda ^2}+\ldots ]\circ 
\]
\[
\circ [1+d\tau \cdot \frac d{d\tau }+\frac 1{2!}\cdot d\tau ^2\cdot \frac{d^2%
}{d\tau ^2}+\ldots ]x^i\}_{(\tau _0,\lambda _0)}= 
\]

\[
=\{[1+k\cdot d\lambda \cdot \frac d{d\lambda }+\frac 12\cdot k^2\cdot
d\lambda ^2\cdot \frac{d^2}{d\lambda ^2}+ 
\]
\[
+d\tau \cdot \frac d{d\tau }+k\cdot d\lambda \cdot d\tau \cdot \frac
d{d\lambda }\circ \frac d{d\tau }+\frac 12\cdot d\tau ^2\cdot \frac{d^2}{%
d\tau ^2}+\cdots ]x^i\}_{(\tau _0,\lambda _0)}\approx 
\]
\[
\approx \{[1+d\lambda \cdot \frac d{d\lambda }+\frac 12\cdot d\lambda
^2\cdot \frac{d^2}{d\lambda ^2}]x^i\}_{(\tau _0,\lambda _0)}\,\,\,\,\,\,\,%
\text{,} 
\]
\[
\{[(k-1)\cdot d\lambda \cdot \frac d{d\lambda }+d\tau \cdot \frac d{d\tau
}+\frac 12\cdot (k^2-1)\cdot d\lambda ^2\cdot \frac{d^2}{d\lambda ^2}+ 
\]
\[
+k\cdot d\lambda \cdot d\tau \cdot \frac d{d\lambda }\circ \frac d{d\tau
}+\frac 12\cdot d\tau ^2\cdot \frac{d^2}{d\tau ^2}]x^i\}_{(\tau _0,\lambda
_0)}\approx 0\,\,\,\,\,\,\text{.} 
\]

Up to the first order of $d\tau $ and $d\lambda $, under the conditions $%
d\lambda <\varepsilon _1\ll 1$, $d\tau <\varepsilon _2\ll 1$, $d\tau \cdot
d\lambda <\varepsilon _1\cdot \varepsilon _2<\varepsilon ^2\ll 1$, we obtain
the relation 
\[
\lbrack (k-1)\cdot d\lambda \cdot \frac{dx^i}{d\lambda }+d\tau \cdot \frac{%
dx^i}{d\tau }]_{(\tau _0,\lambda _0)}\approx 0\,\,\,\,\,\,\,\text{.} 
\]

If we further assume that $d\lambda =\overline{\varepsilon }_1\ll 1$, $d\tau
=\overline{\varepsilon }_2\ll 1$, then 
\begin{eqnarray*}
\lbrack (k-1)\cdot \overline{\varepsilon }_1\cdot \frac{dx^i}{d\lambda }+%
\overline{\varepsilon }_2\cdot \frac{dx^i}{d\tau }]_{(\tau _0,\lambda _0)}
&\approx &0\,\,\,\,\,\text{,} \\
\lbrack (k-1)\cdot \overline{\varepsilon }_1\cdot \xi _{\perp }^i+\overline{%
\varepsilon }_2\cdot u^i]_{(\tau _0,\lambda _0)} &\approx &0\,\,\,\,\,\text{,%
}
\end{eqnarray*}
\[
u_{(\tau _0,\lambda _0)}^i=-\alpha \cdot \xi _{\perp (\tau _0,\lambda _0)}^i%
\text{ \thinspace \thinspace \thinspace ,\thinspace \thinspace \thinspace
\thinspace \thinspace \thinspace \thinspace \thinspace \thinspace \thinspace
\thinspace \thinspace \thinspace \thinspace \thinspace \thinspace \thinspace
\thinspace \thinspace \thinspace \thinspace \thinspace \thinspace \thinspace 
}\alpha =(k-1)\cdot \frac{\overline{\varepsilon }_1}{\overline{\varepsilon }%
_2}\,\,\,\,\text{.} 
\]

The last expressions lead to the relation between the vectors $u$ and $\xi $
at the point $A$: 
\[
u_{(\tau _0,\lambda _0)}=-\alpha \cdot \xi _{\perp (\tau _0,\lambda
_0)}\,\,\,\,\,\,\text{,\thinspace \thinspace \thinspace \thinspace
\thinspace \thinspace \thinspace \thinspace \thinspace \thinspace \thinspace
\thinspace \thinspace }\alpha =(k-1)\cdot \frac{\overline{\varepsilon }_1}{%
\overline{\varepsilon }_2}\,\,\,\,\,\text{.} 
\]

\textit{Remark}. The factor $k$ could be omitted if we chose $(k-1)\cdot 
\overline{\varepsilon }_1=\overline{\varepsilon }_2=\varepsilon $. If $%
\alpha =0$ then $k=1$ and $u_{(\tau _0,\lambda _0)}=0$. The last expression
is in contradiction with the prerequisite for the existence of the two
parametric congruence $x^i=x^i(\tau ,\lambda )$ [$(d\tau /d\lambda )=0$] at
all considered points of $M$. If $u=0$ then $dx^i/d\tau =0$ and $%
x^i=x^i(\lambda )$ at the considered point $A$ with co-ordinates $x^i(\tau
_0,\lambda _0)$.

If the contravariant vector field $u$ is interpreted as the velocity of an
observer then in the vicinity of the intersection point $B$ of the
hypersurface $\tau =\tau _0=const.$ with the hypersurface $\tau =\tau _0+%
\overline{\varepsilon }_2=const.$ the velocity $u$ of the observer will be
proportional to the vector field $\xi _{\perp }$, tangential to the curve $%
x^i=x^i(\tau _0=const.,\lambda )$ and lying in the hypersurface $x^i(\tau
_0=const.,\lambda )$. Since the vector field $\xi _{\perp }$ is orthogonal
to $u$, i.e. since $g(u,\xi _{\perp })=0$, then from the condition $u_{(\tau
_0,\lambda _0)}=-\alpha \cdot \xi _{\perp (\tau _0,\lambda _0)}$, it follows
that 
\begin{eqnarray*}
g(u,\xi _{\perp })_{(\tau _0,\lambda _0)} &=&-\alpha \cdot g(\xi _{\perp
},\xi _{\perp })_{(\tau _0,\lambda _0)}=0\text{ \thinspace \thinspace
\thinspace \thinspace \thinspace \thinspace \thinspace ,\thinspace
\thinspace \thinspace \thinspace \thinspace \thinspace \thinspace \thinspace
\thinspace \thinspace }\alpha \neq 0\text{\thinspace \thinspace \thinspace
\thinspace \thinspace \thinspace ,} \\
g(u,\xi _{\perp })_{(\tau _0,\lambda _0)} &=&-\frac 1\alpha \cdot
g(u,u)=0\,\,\,\,\,\,\text{.}
\end{eqnarray*}

Therefore, the contravariant non-null (non-isotropic) vector fields $u$ and $%
\xi _{\perp }$ [$g(u,u)=e\neq 0$, $g(\xi _{\perp },\xi _{\perp })\neq 0$%
,\thinspace \thinspace \thinspace \thinspace \thinspace $g(u,\xi _{\perp
})=0 $] degenerate to null (isotropic) vector fields [$g(u,u)=e=0$, $g(\xi
_{\perp },\xi _{\perp })=0$,\thinspace \thinspace \thinspace \thinspace
\thinspace $g(u,\xi _{\perp })=0$] in the vicinity of the intersection point 
$B$.

If a pair $(u,\xi _{\perp })$ with $g(u,\xi _{\perp })=0$ of two non-null
vector fields $u$ and $\xi _{\perp }$ degenerate to a pair of null vector
fields in the vicinity of a point belonging to a hypersurface (orthogonal to 
$u$) then this point could be consider as an intersection point of two
hypersurfaces with different constant proper times. The point $A$ could be
consider as a point where the conditions for a frame of reference are
violated and from this point on, in the direction to the intersection point $%
B$, there exists no frame of reference. The point $A$ could be considered as
a boundary point at the hypersurface $x^i(\tau _0=const.,\lambda )$ for
which a frame of reference is not more defined. This fact could be
interpreted physically as the existence of a limited velocity $u$ such that $%
g(u,u)=e=0$ for which an observer could not have a unique determined proper
frame of reference. On the one side, a frame of reference is defined for a
non-null vector field $u$ [$g(u,u)=e\neq 0$] determining its corresponding
projective metrics $h_u=g-(1/e)\cdot g(u)\otimes g(u)$ and $h^u=\overline{g}%
-(1/e)\cdot u\otimes u$ only if $e\neq 0$. On the other side, the existence
of a frame of reference forbids the existence of intersection points of
hypersurfaces, orthogonal to the velocity vector field of the observer. The
last condition leads to the same requirement valid for the existence of
projective metrics for the metric field $u$. Therefore, the degeneration of
a non-null velocity vector field to a null velocity vector field is an
obstacle for the existence of a local frame of reference for the regions and
points where the null vector field is defined.

A hypersurface, orthogonal to the vector $u$, is determined by $n-1$ vectors 
$\xi _{(a)\perp }$, $a=1,...,n-1$, orthogonal to $u$. Every of these vectors
could lie at a curve having intersection point with a curve on an other
hypersurface, orthogonal to $u$ and with different proper time. All
intersection points with their neighborhoods determine the boundaries of the
region where a frame of reference could exist. Therefore, \textit{a local
criteria for the existence of a frame of reference is the existence of a
region in space or space-time where the velocity vector field of an observer
does not degenerate to a null-vector field}.

\section{Conclusion}

In the present paper local conditions are considered for the regions where a
frame of reference could exist. It turned out that there is a simple
condition which could be used as a criteria for the existence of a frame of
reference: A frame of reference could exist for an observer until its
velocity appears as a non-null (non-isotropic) vector field. In general
relativity, this means that the velocity of an observer should be smaller
than the velocity of light in vacuum if the observer wish to see and
describe events around him in the frame of its own frame of reference. If
the velocity of the observer reaches to the velocity of light then he would
be unable to detect and describe the events around him. To our great
surprise, this limitation appears in full correspondence with the conditions
imposed in the definition of a frame of reference on the velocity vector
field to which the corresponding projective metrics belong. It is worth to
be mentioned that the local criteria for the existence of a frame of
reference is valid in all spaces with affine connections and metrics which
could be used as models of space or space-time.

\end{document}